\documentclass{jpp}
\usepackage{graphicx}
\usepackage{hyperref}
\hypersetup{colorlinks = true, linkcolor = blue, citecolor = blue, urlcolor = blue}

\usepackage[utf8]{inputenc}
\usepackage[T1]{fontenc}
\usepackage{amsmath}
\usepackage{tikz}
\usetikzlibrary{decorations.markings}
\usetikzlibrary {arrows.meta}
\usepackage{subcaption}

\newcommand{\vthi}{v_{\textnormal{th}}}
\newcommand{\rhoi}{\rho_i}
\newcommand{\charge}{Ze}
\newcommand{\fzero}{f_0}
\newcommand{\bunit}{\boldsymbol{b}}
\newcommand{\gyroR}[1]{\langle #1 \rangle_{\boldsymbol{R}}}
\newcommand{\gyror}[1]{\langle #1 \rangle_{\boldsymbol{r}}}
\newcommand{\vexb}{\boldsymbol{v}_{\varphi}}
\newcommand{\vdrift}{\boldsymbol{v}_{d}}
\newcommand{\vpa}{v_{\parallel}}
\newcommand{\vpe}{v_{\perp}}
\newcommand{\hk}{\hat{h}_{\boldsymbol{k}}}
\newcommand{\phik}{\hat{\varphi}_{\boldsymbol{k}}}
\newcommand{\bessarg}{\alpha_k}
\newcommand{\jzero}{J_0(\bessarg)}

\newcommand{\omegas}{\omega_*}
\newcommand{\gyroangle}{\vartheta}
\newcommand{\dnk}{\delta N_{\boldsymbol{k}}}
\newcommand{\dpk}{\delta p_{\boldsymbol{k}}}

\newcommand{\gamnfac}{K_1}
\newcommand{\gamTfac}{K_2}
\newcommand{\omegalw}{\omega^{\textnormal{lw}}}

\newcommand{\vpanorm}{\tilde{v}_{\parallel}}
\newcommand{\vpenorm}{\tilde{v}_{\perp}}
\newcommand{\vnorm}{\tilde{v}}
\newcommand{\Tgrad}{L_B/L_T}
\newcommand{\kcut}{k_y^{o}}

\newcommand{\LTeff}{L_T^{\textnormal{eff}}}
\newcommand{\defeq}{\equiv}

\newcommand{\taunl}{\tau_{\textnormal{nl}}}

\usepackage{array}
\shorttitle{Scaling laws for the SWITG cutoff wavenumber}
\shortauthor{O. Gupta, M. Barnes, et al.}

\title{Scaling laws for the cutoff wavenumber of the short-wavelength ion-temperature-gradient mode in a Z-pinch}

\author{O. Gupta\aff{1,2}
  \corresp{\email{om.gupta@maths.ox.ac.uk}},
  M. Barnes\aff{1,3}, F. I. Parra\aff{4}, L. Podavini\aff{5}, A. Zocco\aff{5}, T. Adkins\aff{4} and P. G. Ivanov\aff{6}}

\affiliation{
\aff{1}Rudolf Peierls Centre for Theoretical Physics, University of
Oxford, Oxford OX1 3PU, UK
\aff{2}Mathematical Institute, University of Oxford, OX2 6GG, UK
\aff{3}University College, Oxford OX1 4BH, UK
\aff{4} Princeton Plasma Physics Laboratory, Princeton, NJ 08540, USA
\aff{5} Max-Planck-Institut für Plasmaphysik, 17491 Greifswald, Germany
\aff{6} United Kingdom Atomic Energy Authority, Culham Campus, Abingdon, OX14 3DB, UK
}

\begin{document}

\maketitle

\begin{abstract}
We use a heuristic fluid model to predict the dependence of the cutoff wave number for the short-wavelength ion temperature gradient (SWITG) mode on ion density gradient, ion temperature gradient (ITG) and ion-electron temperature ratio.  In particular, we predict that the cutoff wave number increases linearly with increasing ITG for sufficiently large values of the ITG.  Direct numerical solutions of the gyrokinetic dispersion relation using a purpose-built solver confirm the predicted scalings at large ITG values and find a weaker power-law scaling for intermediate ITG values.  Combining these wave number scalings with a simple diffusive estimate for turbulent fluxes produces a scaling prediction for the ITG heat flux in SWITG-driven turbulence.  Applying the critical balance conjecture additionally provides scalings for the aspect ratio of the SWITG turbulent eddies.

\end{abstract}
\section{Introduction}
Turbulent transport driven by small-scale instabilities is known to limit the energy confinement time in tokamaks~\citep{Yoshida_2025} and optimised stellarators~\citep{Bozhenkov2020}.  Though there is a plethora of such instabilities, each catalogued by the ingredients required, the archetypal instability for ion transport is the curvature-driven ion-temperature-gradient (ITG) instability ~\citep{1968PlPh...10..649P, Terry_1982, 1983PhFl...26..673G, romanelli1989, evensenNF1998}.  It has been the subject of numerous experimental, numerical and analytical studies. While the precise details of the instability depend on the magnetic geometry and the plasma profiles, its key features are a growth rate that increases with increasing ITG beyond a critical value and a characteristic wavelength that is comparable to or larger than the ion Larmor radius.

More recently, it was shown that a short-wavelength version of the ITG instability, referred to as the SWITG mode, exists ~\citep{10.1063/1.859454, 10.1063/1.860537, smolyakovPRL02}.  A key feature of the SWITG identified in previous studies is the appearance of a second maximum in the linear growth rate spectrum at scales smaller than the ion Larmor scale~\citep{pu1985,smolyakovPRL02,hirose2002short,gao2003,gao2005,GAO_SANUKI_ITOH_DONG_2006,chowdhuryPoP09,chowdhuryPoP12,gaj2020,Singh2023,rodríguezJPP2025}.  There are both slab and toroidal variants of the SWITG, driven by resonances with parallel streaming and binormal magnetic drifts, respectively.  The dependence of the SWITG growth rates on various plasma parameters has been explored in some detail in sheared slab and toroidal geometry, with and without electromagnetic effects, and in both local and radially `global' models.  A more limited number of nonlinear simulations have also been conducted.  Due to the complicated nature of many of the models used for these studies, both the range of parameter space considered and the availability of general analytical results have necessarily been limited.

To complement these full-physics studies, we develop here a minimal model to study the curvature-driven variant of the SWITG mode.  This model consists of fluid ions and Boltzmann electrons in a Z-pinch magnetic geometry.  The simplicity of our model enables analytical calculation of mode growth rates and facilitates the identification of a simple scaling for the SWITG mode cutoff scale with both the ion density and temperature gradient scale lengths as well as the ion-electron temperature ratio.  Our scalings are confirmed via numerical solution of the linear dispersion relation with both a purpose-built dispersion solver and with the gyrokinetic code \texttt{stella}~\citep{barnes2019stella}.

The paper is laid out as follows: In Section~\ref{sec:model_system} we describe our model system and state our ordering and other simplifying assumptions, culminating in the linearised gyrokinetic-Poisson equations that we use as the starting point in our analysis.  We then develop a heuristic fluid model in Section~\ref{sec:fluid_model} and evaluate the associated dispersion relation in the long- and short-wavelength limits.  Here the main novel result is an expression for the minimum unstable wavenumber for the SWITG as a function of density and temperature gradient scale lengths, as well as the ion-electron temperature ratio.  The prediction for this cutoff wavenumber is compared with the exact results from a full gyrokinetic analysis in Section~\ref{sec:linGK}.  The possible implications of this scaling for turbulence are explored in Section~\ref{sec:nonlin}, and we conclude with an overview and prospective for future work in Section~\ref{sec:discussion}.

\section{Electrostatic gyrokinetic model in a Z-pinch}\label{sec:model_system}
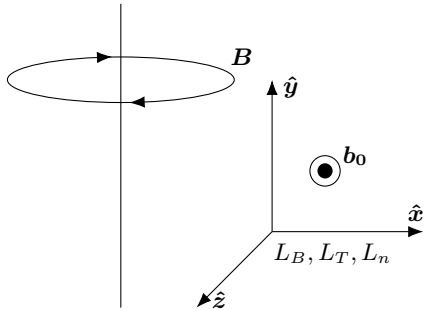
\begin{figure}
    \centering
    \begin{tikzpicture}
        \draw[decoration={markings, mark=at position 0.3 with {\pgftransformscale{1.2}\arrow{Latex[reversed]}}, mark=at position 0.8 with {\pgftransformscale{1.2}\arrow{Latex[reversed]}}},
        postaction={decorate}] (0,0) ellipse (1.5 and 0.3);
        \node at (1.6, 0.3) {$\boldsymbol{B}$};
        \draw (0,1) -- (0,-3);
        \draw[-{Latex[length=2mm]}] (2,-2) -- (4,-2);
        \draw[-{Latex[length=2mm]}] (2,-2) -- (2,-0);
        \draw[-{Latex[length=2mm]}] (2,-2) -- (1,-3);
        \node at (1.3, -2.9) {$\boldsymbol{\hat{z}}$};
        \node at (3.9, -1.75) {$\boldsymbol{\hat{x}}$};
        \node at (2.25, -0.1) {$\boldsymbol{\hat{y}}$};
        \node at (2.8, -2.3) { $L_B, L_T, L_n$};
        \fill[black] (2.7,-1.2) circle (0.1);
        \draw (2.7,-1.2) circle (0.2);
        \node at (3.1, -1) {$\boldsymbol{b_0}$};
        
    \end{tikzpicture}
    \caption{An illustration of the Z-pinch geometry adopted in our analysis. We use orthonormal coordinates $(x,y,z)$ such that a constant curvature magnetic field exists along the $\boldsymbol{\hat{z}}$ direction with all equilibrium gradients taken to be along the $\boldsymbol{\hat{x}}$ direction.}
    \label{fig:geom}
\end{figure}
For simplicity, we restrict our attention to a plasma consisting of electrons of charge $-e$ and a single ion species of mass $m$ and charge $Ze$, immersed in a constant-curvature magnetic field of the form $\boldsymbol{B}=B(x)\hat{\boldsymbol{z}}$.  The equilibrium plasma density $n$ and  temperature $T$ are inhomogeneous, with variation only in the $x$ direction.  Here $(x,y,z)$ is the set of orthonormal coordinates illustrated in Figure~\ref{fig:geom}, with $\hat{\boldsymbol{z}}$ the unit vector in the $z$ direction.  For the case of vanishing magnetic shear considered here, the magnetic geometry is equivalent to that of a Z-pinch, with $x$, $y$ and $z$ the radial, axial and azimuthal coordinates, respectively.

We work in the collisionless, electrostatic limit and assume that electron motion along the magnetic field is sufficiently fast that the electron density is well approximated by a Boltzmann response: $\delta n_e = (e\varphi/T_e)n_e$, where $\delta n_e$ and $n_e$ are the fluctuating and equilibrium components of the electron density, respectively, $\varphi$ is the electrostatic potential and $T_e$ is the equilibrium electron temperature.  Finally, we focus on fluctuations that have no variation along the equilibrium magnetic field: These `flute' perturbations are the most unstable in a constant-curvature system and are the simplest approximation to curvature ITG modes in the bad-curvature region of tokamaks.


\subsection{Gyrokinetic model equations}

We assume the fluctuations in our plasma satisfy the standard $\delta \!f$-gyrokinetic ordering \citep{ant, frieman1982, sugama, Abel_2013}; i.e., fluctuations with characteristic frequency $\omega$ and wavenumbers $k_\parallel$ and $k_\perp$ parallel and perpendicular to the direction of the equilibrium magnetic field $\bunit = \hat{\boldsymbol{z}}$ are ordered such that 
\begin{equation}
    \epsilon \equiv \frac{\rho_i}{L} \sim \frac{\omega}{\Omega} \sim \frac{k_{\parallel}}{k_{\perp}} \sim \frac{e \varphi}{T_{i}} \ll 1,
\end{equation}
where $\Omega = Ze B / m c$ is the ion cyclotron frequency, $\rho_i = \vthi/\Omega$ is the thermal ion Larmor radius, $\vthi = \sqrt{2T_i/m}$ is the ion thermal speed and $L$ is a typical equilibrium length scale.  In addition, it is assumed that all equilibrium quantities evolve on the time scale $\tau^{-1}_E \sim \epsilon^3 \Omega$, so they may be treated as static in the discussion that follows. The fluctuating ion distribution function can then be written to lowest order in $\epsilon$ as
\begin{equation} \label{eq:4}
    \delta f  = -\frac{\charge \varphi}{T_i} \fzero + h, 
\end{equation}
consisting of a Boltzmann part and a gyrokinetic part $h$ that is independent of gyro-phase angle at fixed guiding centre position $\boldsymbol{R}=\boldsymbol{r}-\bunit \times \boldsymbol{v}/\Omega$, with $\boldsymbol{r}$ and $\boldsymbol{v}$ the particle position and velocity, respectively.  To lowest order in $\epsilon$, the equilibrium distribution function $\fzero$ is a Maxwellian,
\begin{equation}
    \fzero = \frac{n_i}{\pi^{3/2}\vthi^3}e^{-v^2/\vthi^2},
\end{equation}
with $n_i=n_i(x)$ the equilibrium ion density and $\vthi=\vthi(x)$.

The resultant (electrostatic, two-dimensional) gyrokinetic evolution equation for $h$ is
\begin{equation}
    \frac{\partial}{\partial t}\left(h - \frac{\charge \gyroR{\varphi}}{T_{i}} \fzero\right) + \vdrift \cdot \nabla_{\perp} h + \vexb\cdot \nabla_{\perp}(h + \fzero) = 0,
\label{eq:gk}
\end{equation}
where $\gyroR{...}$ denotes a gyro-average at constant guiding centre position and $\vexb$ and $\vdrift$ are the (gyro-averaged) $E\times B$ and magnetic drift velocities, respectively.  In an arbitrary geometry with vanishing plasma beta they are given by
\begin{equation}
    \vexb = \frac{c}{B} \bunit \times \nabla \gyroR{\varphi}, \quad \vdrift = \frac{\bunit}{\Omega} \times \nabla \ln{B}\left(v^2_\parallel + \frac{v^2_\perp}{2}\right),
\end{equation}
with $\vpa$ and $\vpe$ the components of $\boldsymbol{v}$ along and across $\bunit$, respectively.

The gyrokinetic system is closed by enforcing quasineutrality,
\begin{equation}
    \delta n_e = Z \delta n_i = Z\int \mathrm{d}^3\boldsymbol{v} \left(-\frac{\charge \varphi}{T_{i}} n_i + \int \mathrm{d}^3 \boldsymbol{v} \; \gyror{h} \right), \label{eq:1}
\end{equation}
where $\gyror{...}$ denotes an average over gyro-angle at fixed $\boldsymbol{r}$.  Upon application of the Boltzmann electron assumption, \eqref{eq:1} can be rearranged to provide an explicit equation for $\varphi$ in terms of $h$:
\begin{equation}
    \frac{e\varphi}{T_e}\left(1+\tau^{-1}\right)n_i = \int \mathrm{d}^3\boldsymbol{v}\gyror{h},
\label{eq:qn}
\end{equation}
where we have defined $\tau\equiv T_i/ZT_e$.

\subsection{Fourier representation}

In our linear analysis of~\eqref{eq:gk} and~\eqref{eq:qn}, we will find it convenient to work within a Fourier representation in both space and time. In deriving the gyrokinetic equation \eqref{eq:gk}, we have assumed equilibrium quantities and their local gradients can be taken as constant on the space-time scales of the fluctuations. Hence, all coefficients multiplying $h$ and $\varphi$ in~\eqref{eq:gk} and~\eqref{eq:qn} are independent of $x$, $y$, $z$ and $t$.

We decompose $h$ and $\varphi$ in the Fourier series
\begin{equation}
    h =  \sum_{\boldsymbol{k}} \hat{h}_{\boldsymbol{k}} (\vpa,\vpe,t)e^{i \boldsymbol{k} \cdot \boldsymbol{R}}, \quad \quad \varphi =  \sum_{\boldsymbol{k}} \hat{\varphi}_{\boldsymbol{k}}(t) e^{i \boldsymbol{k} \cdot \boldsymbol{r}}.
\label{eq:3}
\end{equation}
Neglecting the nonlinear term in~\eqref{eq:gk} and performing a Fourier transform in time results in the linearised gyrokinetic equation for the Fourier coefficient $\hk$,
\begin{equation}
\begin{split}
    \omega &\left(\hk - \frac{\jzero}{\tau}\frac{e\phik}{T_e}\fzero\right) - \omega_d \left(\frac{2\vpa^2}{\vthi^2}+\frac{\vpe^2}{\vthi^2}\right)\hk \\
    & + \omegas \left[1 + \eta\left(\frac{v^2}{\vthi^2}-\frac{3}{2}\right)\right]\frac{\jzero}{\tau}\frac{e\phik}{T_e}\fzero = 0,
\label{eq:gkk}
\end{split}
\end{equation}
where $J_0$ is the zeroth-order Bessel function of the first kind, $\bessarg=k_{\perp}v_{\perp}/\Omega$, $\eta=L_n/L_T$, $L^{-1}_X=-d\ln X/dx$, with $X$ one of $B$, $n$ or $T$, and the drift and diamagnetic frequencies are, respectively,
\begin{equation}
    \omega_d = - \frac{k_y\rhoi}{2} \frac{\vthi}{L_B} \quad \quad 
    \omegas = - \frac{k_y\rhoi}{2}\frac{\vthi}{L_n}.
\end{equation} 
To evaluate the gyro-averages in Fourier space, we have made use of the identity~\citep{abramowitz1968handbook}
\begin{equation}
    \gyroR{\varphi (\boldsymbol{r}, t)} = \sum_{\boldsymbol{k}}\phik e^{i \boldsymbol{k}\cdot \boldsymbol{R}} \frac{1}{2\pi} \oint \mathrm{d}\gyroangle \; \exp \left(i \frac{k_{\perp} v_{\perp}}{\Omega} \cos(\gyroangle)\right) = \sum_{\boldsymbol{k}}J_0\left(\bessarg\right) \phik e^{i \boldsymbol{k} \cdot \boldsymbol{R}},
\end{equation}
where $\gyroangle$ is the gyro-angle.

Similarly, the Fourier representation of the quasineutrality condition~\eqref{eq:qn} is
\begin{equation}
    \frac{e\phik}{T_e}\left(1+\tau^{-1}\right)n_i  = \int \mathrm{d}^3\boldsymbol{v} \ J_0(\bessarg)\hk,
\label{eq:qnk}
\end{equation}
where 
\begin{equation}
    \gyror{h(\boldsymbol{R}, \vpe, \vpa, t)} = \sum_{\boldsymbol{k}}\hk e^{i \boldsymbol{k} \cdot \boldsymbol{r}} \frac{1}{2\pi} \oint \mathrm{d}\gyroangle \; \exp \left(-i \frac{k_{\perp} v_{\perp}}{\Omega} \cos(\gyroangle)\right) = \sum_{\boldsymbol{k}}J_0\left(\bessarg\right) \hk e^{i \boldsymbol{k} \cdot \boldsymbol{r}}.
\end{equation}
Equations~\eqref{eq:gkk} and~\eqref{eq:qnk} form a closed system of equations for the Fourier coefficients $\hk$ and $\phik$.

\section{Heuristic fluid model}
\label{sec:fluid_model}

Before undertaking a direct gyrokinetic analysis, we first construct and examine a heuristic fluid model.  We will see that a quantitatively correct treatment of the short-wavelength ITG instability requires retention of resonant effects that are absent from this simple fluid model.  Nonetheless, the fluid model captures qualitatively the salient features of the instability and facilitates a comparison with the standard long-wavelength ITG instability, including its stabilisation by a density gradient.

We take as our starting point the gyrokinetic equation~\eqref{eq:gkk} and the quasineutrality condition~\eqref{eq:qnk} in Fourier space.  Multiplying
\eqref{eq:gkk} by $J_0(\bessarg)/n_i$ and integrating over velocity space gives
\begin{equation}
    \omega \left[\frac{\dnk}{n_i} - \frac{\Gamma_0(b_k)}{\tau}\frac{e\phik}{T_e}\right] - \omega_d \frac{\dpk}{n_iT_i} + \left[\Gamma_0(b_k) + \eta b_k \left(\Gamma_1(b_k)-\Gamma_0(b_k)\right)\right] \frac{\omega_*}{\tau} \frac{e\phik}{T_e} = 0,
\label{eq:nk}
\end{equation}
where 
\begin{equation}
    \dnk \equiv \int d^3 \boldsymbol{v} \ J_0(\bessarg) \hk
\end{equation}
is the density deviation from an ion Boltzmann response,
\begin{equation}
   \dpk \equiv\int d^3 \boldsymbol{v} \ \frac{m}{2}\left(2v_{\parallel}^2 + v_{\perp}^2\right) J_0(\bessarg)\hk
\end{equation}
is a combination of the parallel and perpendicular fluctuating pressures, $b_k=k_{\perp}^2\rhoi^2/2$ and $\Gamma_{j}(b)=\exp(-b)I_{j}(b)$, with $I_{j}$ the $j$th modified Bessel function of the first kind.

We close our fluid model by multiplying~\eqref{eq:gkk} by $[J_0(\bessarg)/n_i](2\vpa^2+\vpe^2)/\vthi^2$, integrating over the velocity space, and neglecting the term proportional to $\omega_d$.  This is an \textit{ad hoc} assumption that can be justified if $\omega \gg \omega_d$ or if the moments of $h$ beyond the energy moment are asymptotically small.  For long-wavelength ITG modes, the solutions for $\omega$ thus obtained satisfy $\omega\gg\omega_d$, while for short-wavelength ITG modes they do not.  This will lead us to conduct a full gyrokinetic analysis of the short-wavelength instability in the next section, but we will find that the correct qualitative features of the instability are captured by the fluid approach.  The resulting equation for $\dpk$ is
\begin{equation}
    \omega \left[\frac{\dpk}{n_iT_i} - \frac{\gamnfac(b_k)}{\tau} \frac{e\phik}{T_e}\right] + \left[\gamnfac(b_k)+\eta\gamTfac(b_k)\right]\frac{\omega_*}{\tau} \frac{e\phik}{T_e} = 0,
\label{eq:Tk}
\end{equation}
where $\gamnfac$ and $\gamTfac$ are defined to be
\begin{equation}
    \gamnfac(b_k)=\left(2-b_k\right)\Gamma_0(b_k)+b_k\Gamma_1(b_k)
\end{equation}
and
\begin{equation}
   \gamTfac(b_k) =2\left(b_k-1\right)^2\Gamma_0(b_k) + b_k\left(3-2b_k\right)\Gamma_1(b_k).
\end{equation}

Combining the density equation~\eqref{eq:nk}, the
pressure equation~\eqref{eq:Tk} and the quasineutrality condition~\eqref{eq:qnk} results in a quadratic dispersion relation for $\omega$:
\begin{equation}
    a_2\omega^2 + a_1 \omega + a_0 = 0,
\end{equation}
where $a_0=\omega_d\omega_*(\gamnfac+\eta \gamTfac)$, $a_1=-\omega_d\gamnfac+\omega_*[\Gamma_0+\eta b_k(\Gamma_1-\Gamma_0)]$ and $a_2=\tau+1 -\Gamma_0$.  The solutions for $\omega$ are then
\begin{equation}
    \omega_{\pm} = -\frac{a_1}{2a_2}\pm\frac{\sqrt{a_1^2-4a_0a_2}}{2a_2}.
\label{eq:fluidomega}
\end{equation}

\subsection{Long-wavelength limit}
\label{sec:long_wavelength}

Evaluating~\eqref{eq:fluidomega} for $\Tgrad \gg 1$ and $\eta \sim \tau \sim 1$ in the long-wavelength limit $b_k\ll1$, we obtain
\begin{equation}
    \omegalw_{\pm} \approx -\frac{\omega_*}{2\tau}\pm \frac{\sqrt{\omega_*^2-8\omega_*\omega_d\tau\left(1+\eta\right)}}{2\tau},
\label{eq:omegaLW}
\end{equation}
where we have used $\Gamma_0(b_k)\approx 1$ and $\Gamma_1(b_k)\approx 0$.
When the temperature gradient is much steeper than the density gradient so that $\eta\omega_d\tau\gg \omega_*$, we recover the standard (fluid) growth rate for strongly-driven, long-wavelength curvature-driven ITG modes~\citep{hortonPoF1981}:
\begin{equation}
    \omegalw_{\pm} \approx\pm \mathrm{i}k_y\rhoi\frac{c_s}{\sqrt{L_BL_p}},
\end{equation}
with $c_s=\sqrt{T_e/m}$ the sound speed and $L_p^{-1}=L_{n}^{-1}+L_T^{-1}$ the inverse ion pressure gradient scale length.  However, if $\eta \omega_d \tau\sim \omega_*$, then the term under the radical in~\eqref{eq:omegaLW} can vanish.  At this marginal stability point, we find a critical $\eta$ below which the long-wavelength ITG mode is stabilised:
\begin{equation}
    \eta_{\textnormal{crit}}\approx\sqrt{\frac{1}{8\tau}\frac{L_B}{L_T}}. \label{eq:etacrit}
\end{equation}
Note that for $\eta \sim \eta_{\textnormal{crit}} \gg 1$, the associated frequency is $\omega = -\omega_*/2\tau \gg \omega_d$, consistent with the assumption we used to close our fluid model.

\subsection{Short-wavelength limit}
\label{sec:short_wavelength}

Next, we evaluate~\eqref{eq:fluidomega} in the short-wavelength limit, for which
\begin{equation}
    \Gamma_0(b_k)\approx \frac{1}{\sqrt{\pi}}\frac{1}{k_{\perp}\rhoi}\left(1+\frac{1}{4}\frac{1}{k_\perp^2\rhoi^2}\right), \quad \Gamma_1(b_k)\approx \frac{1}{\sqrt{\pi}}\frac{1}{k_{\perp}\rhoi}\left(1-\frac{3}{4}\frac{1}{k_{\perp}^2\rhoi^2}\right).
\end{equation}
Inserting these expansions into~\eqref{eq:fluidomega} and taking $\Tgrad \gg 1$ gives
\begin{equation}
    \gamnfac \approx \frac{3}{2\sqrt{\pi}}\frac{1}{k_{\perp}\rhoi}, \quad \gamTfac \approx\frac{3}{4\sqrt{\pi}}\frac{1}{k_{\perp}\rhoi},
\end{equation}
\begin{equation}
    a_0 \approx \omega_d\omega_*\left(2+\eta\right)\frac{3}{4\sqrt{\pi}}\frac{1}{k_{\perp}\rhoi},
\end{equation}
\begin{equation}
    a_1 \approx\frac{1}{2\sqrt{\pi}}\frac{1}{k_{\perp}\rhoi}\left[\left(2-\eta\right) \omega_* - 3\omega_d\right],
\end{equation}
and
\begin{equation}
    a_2\approx 1+\tau - \frac{1}{\sqrt{\pi}} \frac{1}{k_{\perp}\rho_i}.
\end{equation}
For fixed values of $\eta$, $L_B/L_n$ and $\tau$, marginal stability occurs at the cutoff wave number $k_{\perp}^o$, given by
\begin{equation}
    k^o_{\perp}\rhoi = \frac{(2-\eta)^2}{\left(2+\eta\right)}\frac{L_B}{12 L_n^{\textnormal{eff}}}, \label{eq:outerscale}
\end{equation}
where we have defined the effective density gradient scale length $L_n^{\textnormal{eff}}\defeq(1+\tau)L_n$.  Note that for $\eta\rightarrow \infty$, this cutoff wavelength simplifies to $k_{\perp}^o\rhoi = (1/12)(L_B/L_T^{\textnormal{eff}})$, with $L_T^{\textnormal{eff}}\defeq (1+\tau)L_T$.

\section{Linear gyrokinetic analysis}
\label{sec:linGK}

To capture correctly the short-wavelength instability with $\omega \sim \omega_d$, we return to the gyrokinetic system given by Eqs.~\eqref{eq:gkk} and~\eqref{eq:qnk}.  Rearrangement of~\eqref{eq:gkk} provides an expression for the non-Boltzmann part of the fluctuating distribution function $\hk$:
\begin{equation}
    \hk = \frac{\omega - \omega_*\left[1+\eta\left(\vnorm^2-3/2\right)\right]}{\omega - \omega_d\left(2\vpanorm^2 + \vpenorm^2\right)} \frac{J_0(\bessarg)}{\tau}\frac{e\phik}{T_e} f_0,
\label{eq:hk}
\end{equation}
where the tildes on the velocity variables indicate normalisation by $\vthi$.  Substitution of~\eqref{eq:hk} into the quasineutrality condition~\eqref{eq:qnk} results in the dispersion relation
\begin{equation}
    1 + \tau = \frac{2}{\sqrt{\pi} }\int_{-\infty}^{\infty} \mathrm{d} \vpanorm\int_0^{\infty} \mathrm{d} \vpenorm \vpenorm \frac{\omega - \omega_*\left[1+\eta\left(\vnorm^2-3/2\right)\right]}{ \omega -\omega_d\left(2\vpanorm^2+\vpenorm^2\right)} e^{-\vnorm^2} J^2_0(\bessarg).
\label{eq:gk_dispersion}
\end{equation}

Our program in this section consists of solving Eq.~\eqref{eq:gk_dispersion} for $\omega$ across a range of values for $k_{\perp} \rho_i$, $\eta$, $\Tgrad$ and $\tau$, with a particular emphasis on the short-wavelength, steep gradient regime.  The solutions we obtain are semi-analytical and are facilitated by using a convergent power series expansion for $J_0^2(\bessarg)$~\citep{Ivanov_Adkins_2023}, as described in Appendix~\ref{app:flr_solver}.  This approach is computationally efficient: Calculating $\omega$ for a single $\boldsymbol{k_{\perp}}$ takes less than a second on a single core of a home laptop.  Our solutions are also verified against data from linear gyrokinetic simulations obtained using the flux tube gyrokinetic code \texttt{stella}~\citep{barnes2019stella}.
Unless otherwise specified, all \texttt{stella} simulations were run with 48 grid points in parallel velocity and 12 in magnetic moment, with maximum values in both $\vpa$ and $v_{\perp}$ of three times the ion thermal speed.


As there is no radial component of the magnetic drift in a Z-pinch, the wave number $k_x$ appears only via $k_\perp$ when it appears as an argument to various Bessel functions.  Although we consider the impact of including $k_x$ at the end of this section, most of our discussion involves setting $k_x = 0$, $k_{\perp} \equiv k_y$ and analysing 1D spectral plots. We also set $\tau = 1$ unless otherwise specified.

\subsection{$\eta \rightarrow \infty$ limit}

\begin{figure}
  \centering
  
  \subcaptionbox{}[0.52\linewidth]{%
    \includegraphics[width=\linewidth]{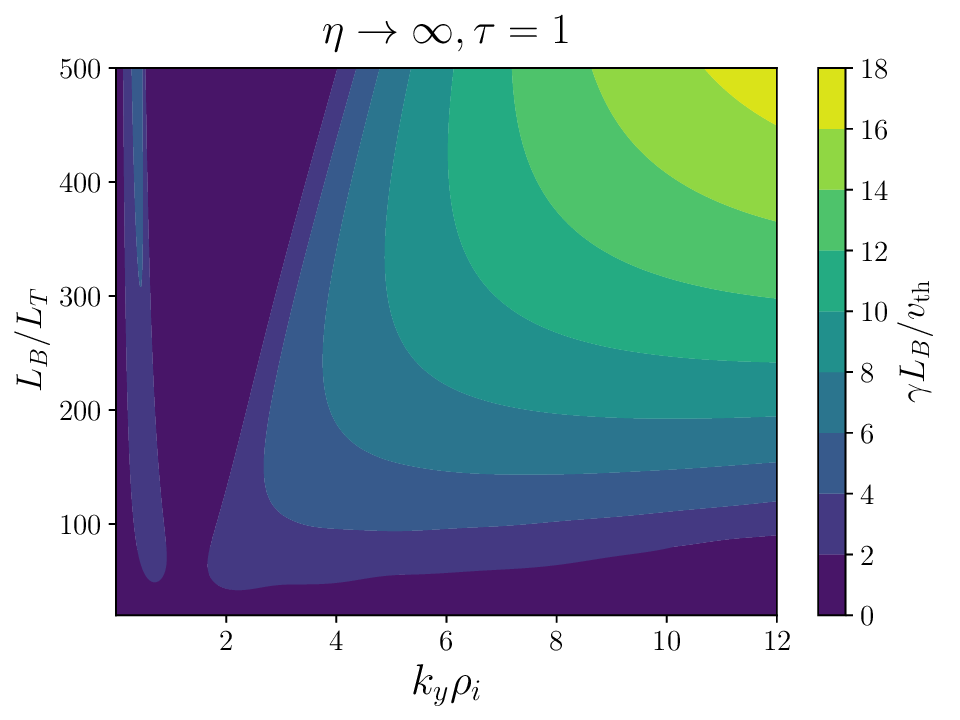}%
  }
  \subcaptionbox{}[0.52\linewidth]{%
    \includegraphics[width=\linewidth]{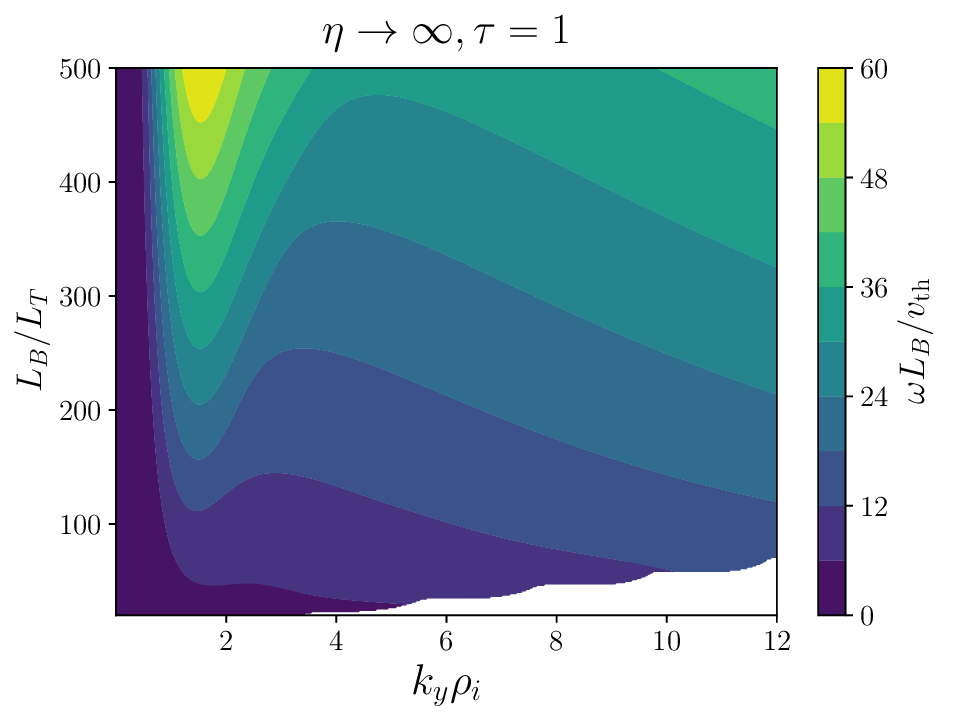}%
  }

  \caption{2-D scan in $\Tgrad$ and $k_y \rho_i$ of the (a) growth rate and (b) real frequency obtained with the solver for $\tau = 1$ and no density gradient ($\eta \to \infty$). Note the presence of two distinct peaks in the growth rate spectrum and a non-monotonic frequency spectrum, both key features of the SWITG. The real frequency is not shown in a narrow region at small drive and very short-wavelength where no growing modes exist and it is numerically challenging to calculate.}
  \label{fig:eta_infty_2D_growthrate}
\end{figure}

\begin{figure}
  \centering

  \subcaptionbox{}[0.51\linewidth]{%
    \includegraphics[width=\linewidth]{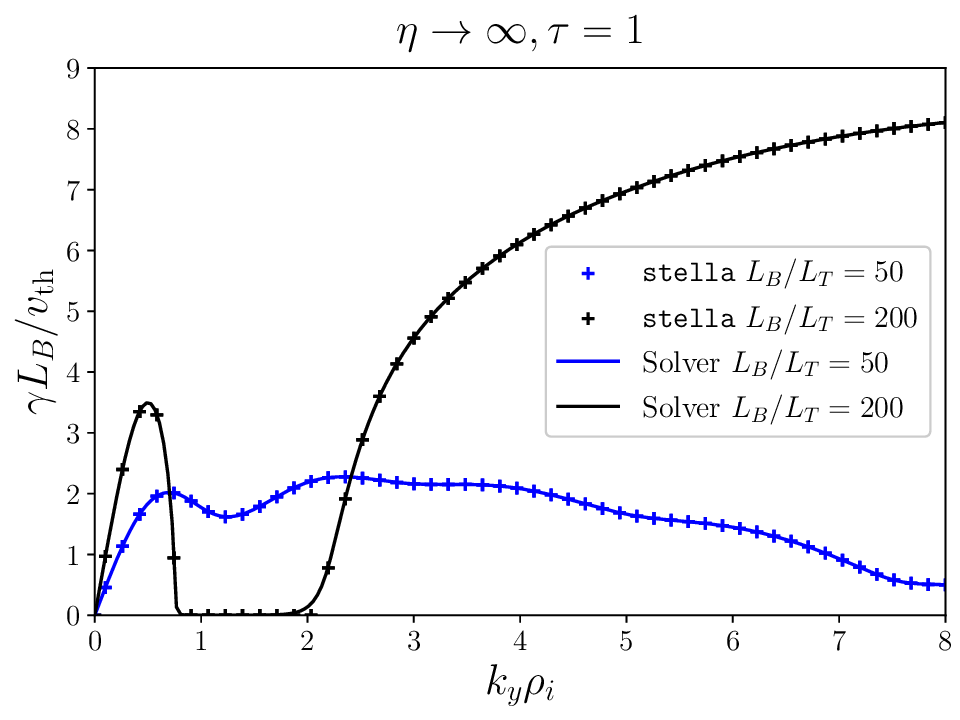}%
  }%
  \subcaptionbox{}[0.51\linewidth]{%
    \includegraphics[width=\linewidth]{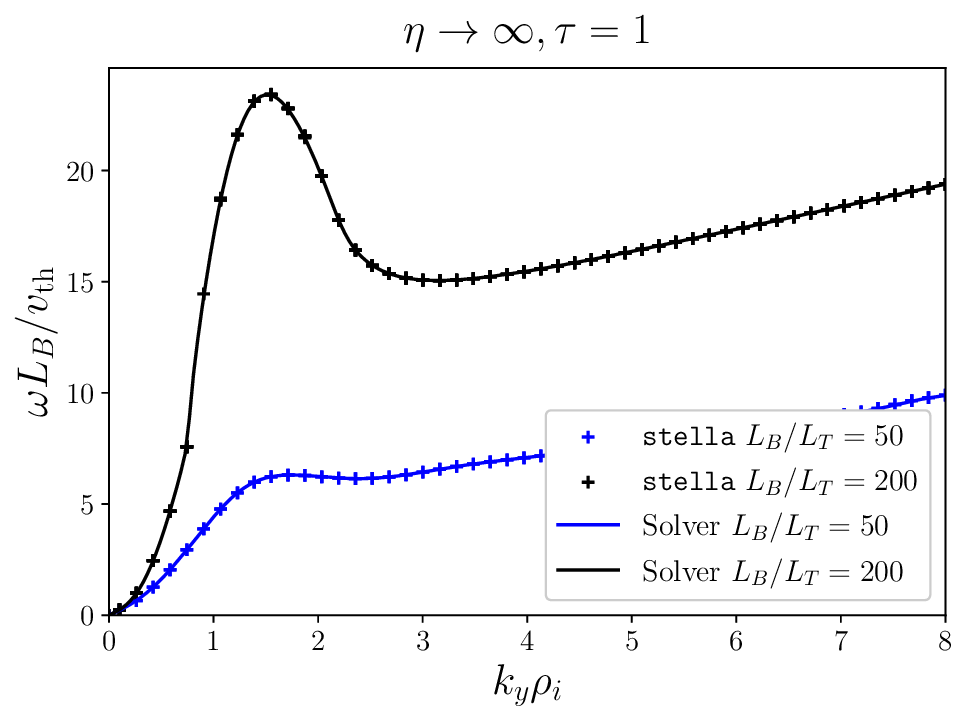}%
  }

  \caption{Plots of the (a) growth rate and (b) real frequency spectra for $\Tgrad=50$ and $200$ with $\tau=1$ and $\eta \to \infty$. The growth-rate spectra show both the long-wavelength ITG mode and the short-wavelength SWITG branch driven by FLR effects. Despite the two instability regions, the real frequency remains continuous and always in the ion diamagnetic direction.}
  \label{fig:eta_infty_1Dcut_growthrate}
\end{figure}

We first consider the case with no equilibrium density gradient, for which $\eta\rightarrow\infty$.  The dependence of the growth rate and real frequency on $k_y \rhoi$ and $\Tgrad$ for this case is given in Fig.~\ref{fig:eta_infty_2D_growthrate}. The long-wavelength ITG instability discussed in Section~\ref{sec:long_wavelength} is present when $k_y\rhoi \lesssim 1$, and the short-wavelength ITG instability discussed in Section~\ref{sec:short_wavelength} is present when $k_y\rhoi \gtrsim 1$.  The instability range for these modes overlaps for small enough $\Tgrad$, but a gap develops between them as $\Tgrad$ increases.  This is seen more clearly in Fig.~\ref{fig:eta_infty_1Dcut_growthrate}, where one-dimensional cuts of the growth rate and real frequency are taken at two different values of $\Tgrad$ (with additional examples provided in Appendix~\ref{app:B}).  As anticipated in Sec.~\ref{sec:long_wavelength}, the long-wavelength ITG growth rate is linearly proportional to $k_y \rho_i$ for $k_y\rhoi \ll 1$ and is suppressed when $k_y\rhoi \gtrsim1$.  In contrast, the short-wavelength ITG mode requires a minimum $k_y$ for instability, and this minimum value increases with $\Tgrad$.  This leads to the unusual behaviour of the growth rate decreasing with $\Tgrad$ over a range of $k_y$ values ($1 \lesssim k_y\rhoi \lesssim 3$ for the $\Tgrad$ values considered in Fig.~\ref{fig:eta_infty_1Dcut_growthrate}).

\subsection{$\eta \sim 1$ limit}

\begin{figure}
  \centering
  \subcaptionbox{}[0.51\linewidth]{%
    \includegraphics[width=\linewidth]{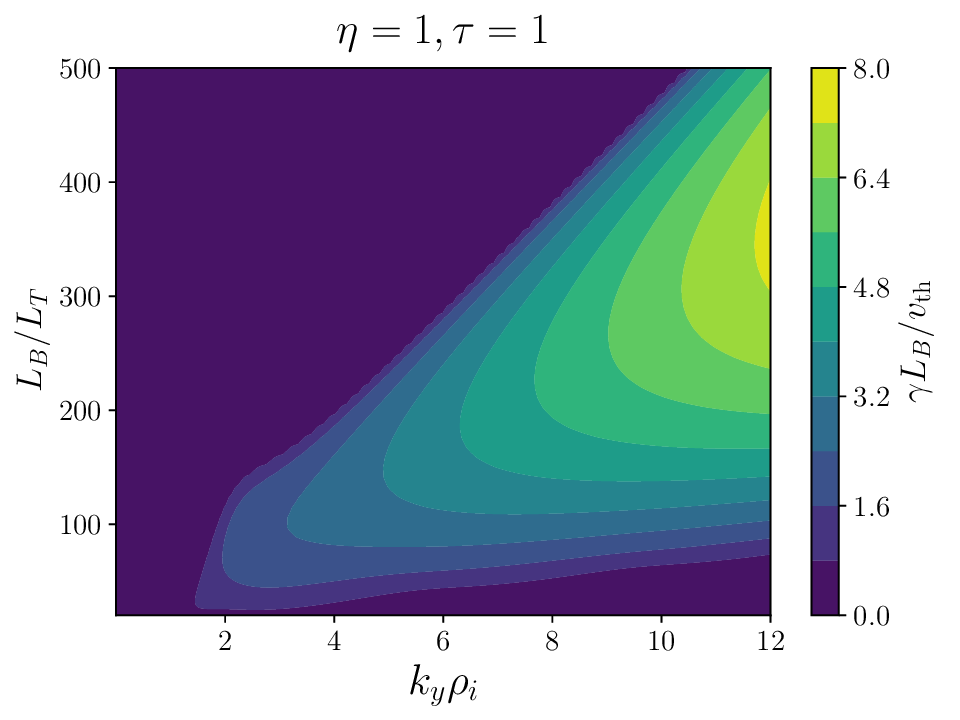}%
  }%
  \subcaptionbox{}[0.51\linewidth]{%
    \includegraphics[width=\linewidth]{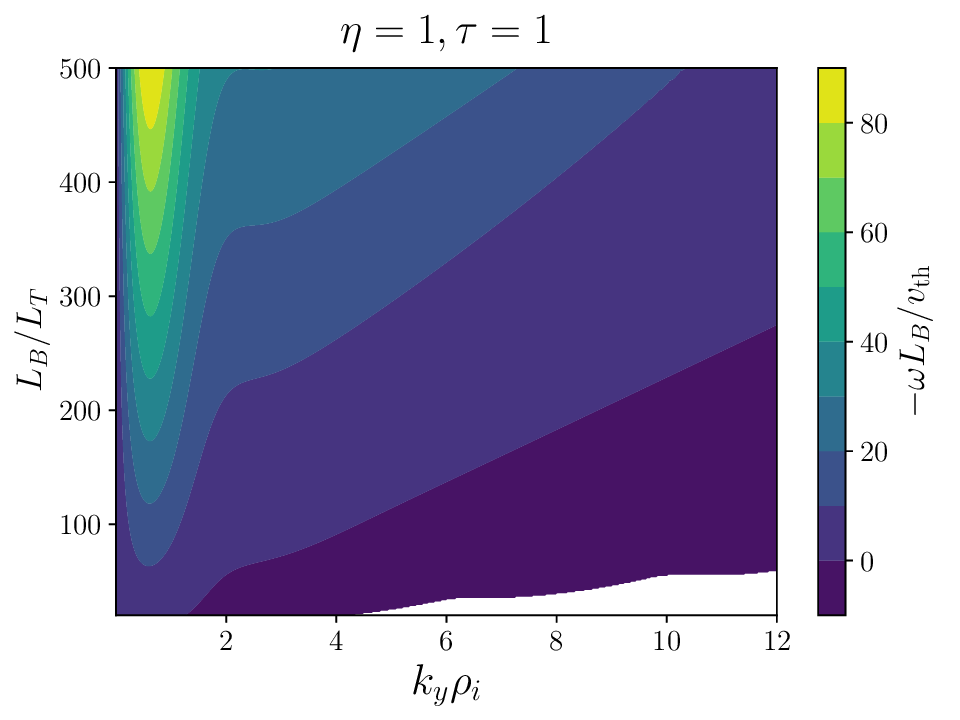}%
  }

  \caption{2-D scan in $\Tgrad$ and $k_y \rho_i$ of the (a) growth rate and (b) real frequency obtained with the solver for $\tau = \eta =1$. We note that the cutoff wave numbers where the growth rates pass through zero increase with $\Tgrad$. Note that we plot the negative of the real frequencies here, so that the modes are propagating in the electron diamagnetic direction in contrast to the $\eta\to\infty$ case shown in Fig.~\ref{fig:eta_infty_2D_growthrate}.}
  \label{fig:eta1_2D_growthrate}
\end{figure}

\begin{figure}
  \centering
  \subcaptionbox{}[0.51\linewidth]{%
    \includegraphics[width=\linewidth]{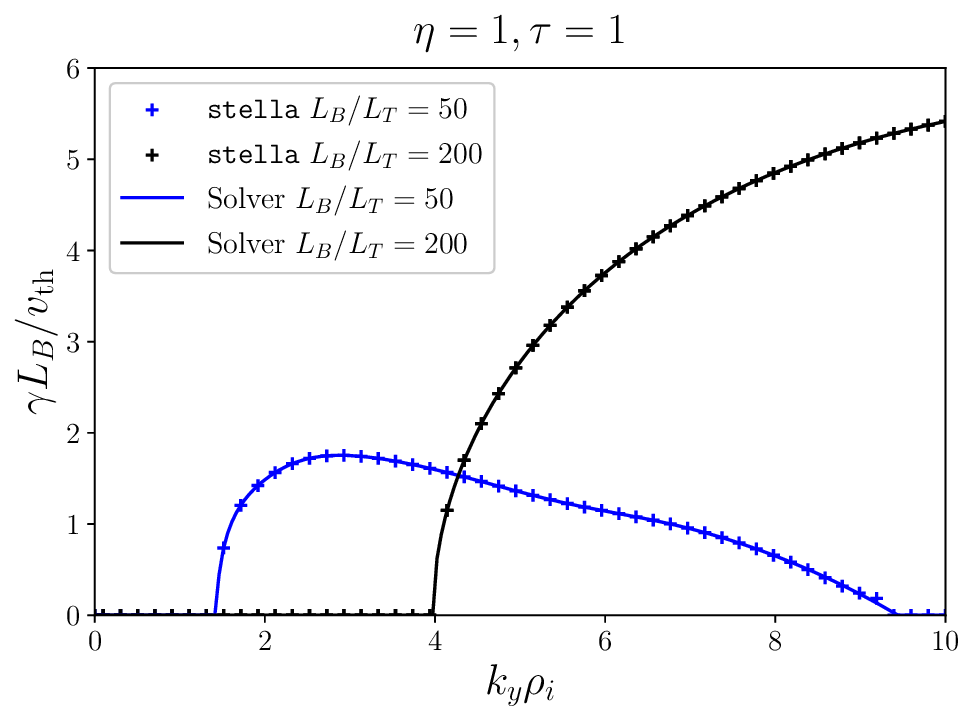}%
  }%
  \subcaptionbox{}[0.51\linewidth]{%
    \includegraphics[width=\linewidth]{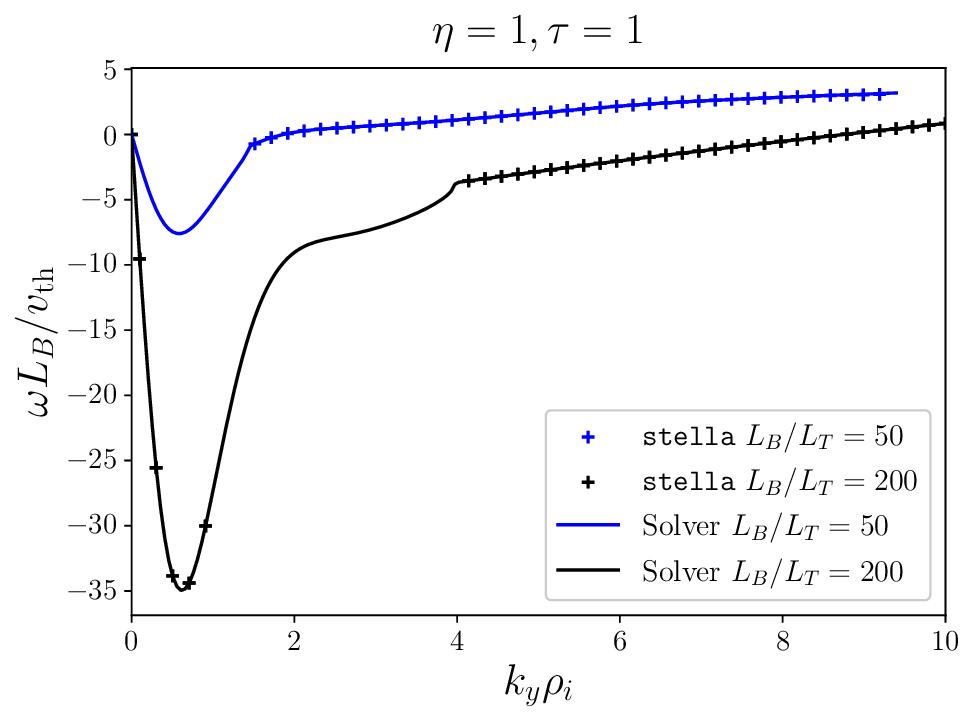}%
  }
  \caption{Plots of the (a) growth rate  and (b) real frequency spectra for $\Tgrad = 50$ and $200$ with $\tau=\eta = 1$. As in the case where the density gradient is absent, we note that the cutoff wave number increases with $\Tgrad$. The \texttt{stella} simulations used for this plot were run with 80 grid points in parallel velocity and 28 grid points in magnetic moment. No \texttt{stella} data points are plotted for the real frequency over part of the $k_y\rhoi$ range due to the difficulty in obtaining the frequency for damped modes.}
  \label{fig:eta1_1Dcut_growthrates}
\end{figure}

We now consider the addition of an equilibrium density gradient of magnitude comparable to the temperature gradient, i.e., $\eta\sim1$. The growth rates and real frequencies for a range of $k_y\rhoi$ and $\Tgrad$ values are plotted in Fig.~\ref{fig:eta1_2D_growthrate} when $\eta=1$.  As observed in previous studies and anticipated in Sec.~\ref{sec:long_wavelength}, the long-wavelength ITG mode is eliminated for $\eta$ below some critical value $\eta_{\textnormal{crit}}$.  It is possible to quantify this relation further through a scan of $\eta_{\mathrm{crit}}$ against $\Tgrad$, as in Fig.~\ref{fig:criteta} (see Appendix~\ref{app:B} for additional plots of the growth rate spectra at different $\eta$ values). We find that the prediction for the critical $\eta$ given by Eq.~\eqref{eq:etacrit} is a good approximation for the long-wavelength ITG. For $\eta < \eta_{\mathrm{crit}}$ the overall stability of the system is determined by the short-wavelength ITG mode.  This is particularly clear in the one-dimensional cuts at two different $\Tgrad$ values shown in Fig.~\ref{fig:eta1_1Dcut_growthrates}.

\begin{figure}
    \centering
    \includegraphics[width=0.8\linewidth]{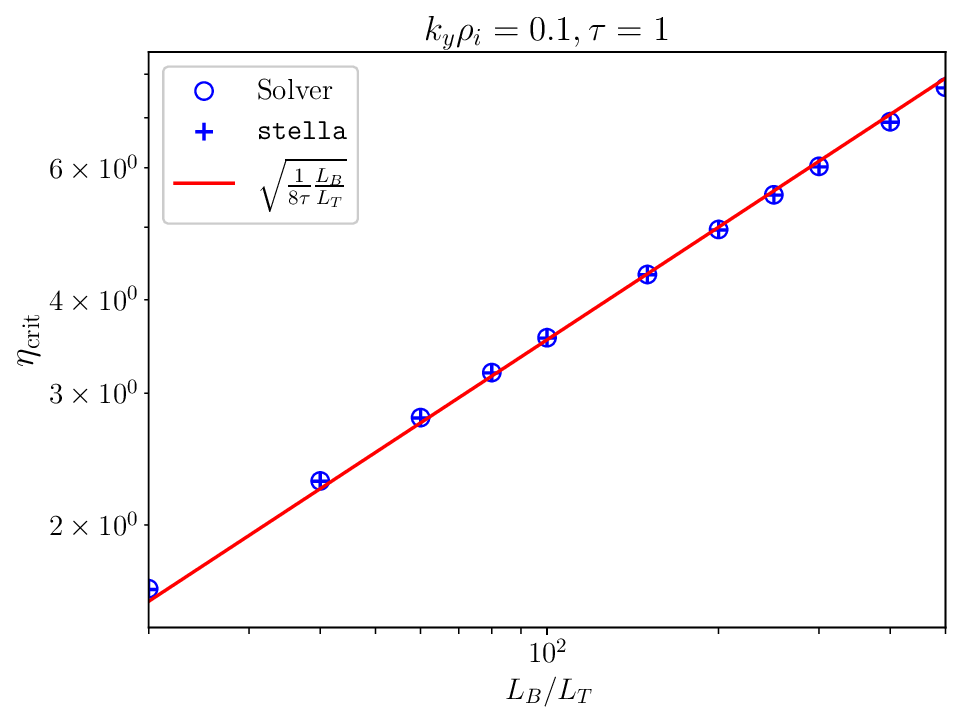}
    \caption{Plot of $\eta_{\mathrm{crit}}$ against $\Tgrad$ obtained from the semi-analytical solver (blue circles), the gyrokinetic code \texttt{stella}, (blue crosses), and the analytical prediction of Eq.~\eqref{eq:etacrit} at $k_y \rho_i = 0.1$.}
    \label{fig:criteta}
\end{figure}

\begin{figure} 
    \centering
    \includegraphics[width=0.95\linewidth]{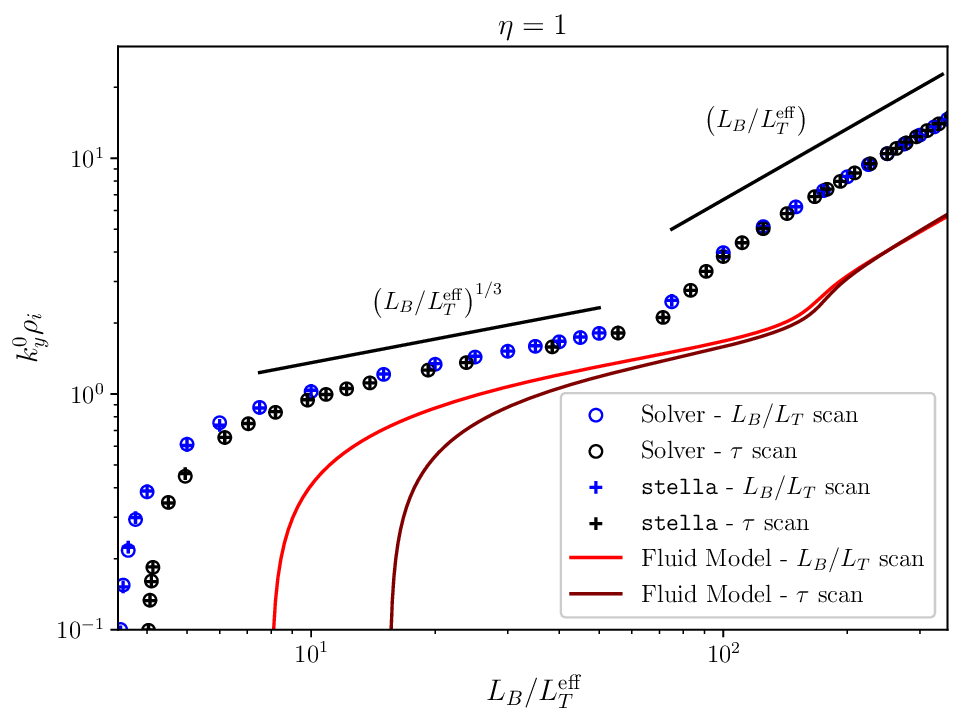}
    \caption{Plot of the cutoff wave number $k_y^o \rho_i$ against $L_B/L_T^{\textnormal{eff}}$ at $\eta = 1$, obtained from the semi-analytical solver (circles), the gyrokinetic code \texttt{stella} (crosses), and the heuristic fluid model from Sec.~\ref{sec:short_wavelength} (red lines).  Blue markers represent data obtained through a scan in $\Tgrad$ at $\tau = 1$ where $\Tgrad \in [6.9,700]$, and black markers represent data obtained through a scan in $\tau$ at $\Tgrad = 500$ where $\tau \in [0.4,123]$. We identify three distinct regimes: the drift kinetic limit for $k_y^o \rho_i \ll 1$; the intermediate regime around $k_y^o\rhoi\sim1$, where the cutoff wave number scales as the power law $(L_B/L_T^{\textnormal{eff}})^{1/3}$; and the $k_y^o \rhoi \gg 1$ limit, where $k_y^o$ scales as $L_B/L_T^{\textnormal{eff}}$. Note that the fluid results for both scans asymptote to the prediction in Eq.~\ref{eq:outerscale}.} \label{fig:eta1_kcut_scaling}
\end{figure}

It is evident from both Figs.~\ref{fig:eta1_2D_growthrate} and~\ref{fig:eta1_1Dcut_growthrates} that there is a minimum wave number required for the short-wavelength mode to be unstable and that this cutoff wave number increases with $\Tgrad$.  We plot this cutoff wave number, $\kcut \rho_i$, against $L_B/L_T^{\textnormal{eff}}$ (for $\tau = 1$) in Fig.~\ref{fig:eta1_kcut_scaling}.  For relatively small values of $\Tgrad$ ($\Tgrad \lesssim10$), the cutoff wave number satisfies $\kcut\rhoi \ll 1$ and increases rapidly with $\Tgrad$.  At moderate $\Tgrad$ values ($10\lesssim\Tgrad\lesssim100$), $\kcut\rhoi\sim 1$ and has an apparent power-law scaling of $\kcut\rhoi\propto(\Tgrad)^{1/3}$.  As $\Tgrad$ increases further, $\kcut\rhoi$ becomes large: in this asymptotic limit, $\kcut\rhoi$ scales like $\Tgrad$, consistent with the prediction of our heuristic fluid model developed in Sec.~\ref{sec:short_wavelength}.  In fact, the fluid model does a remarkably good job of predicting the variation of $\kcut \rho_i$ with $\Tgrad$ across the full range of $\Tgrad$ considered, albeit with an approximately constant offset in $\Tgrad$.  This is despite the mode satisfying the condition $\omega\sim\omega_d$, which is inconsistent with the assumption $\omega \gg \omega_d$ underlying the fluid model.

In the limit of large $\Tgrad$, the term proportional to $\omega$ in the numerator of the gyrokinetic dispersion Eq. relation~\eqref{eq:gk_dispersion} is small. The scaling of $\kcut \rho_i$ with $L_B/\LTeff$, the effective temperature gradient defined below Eq.~\eqref{eq:outerscale}, is then the same (up to a re-scaling) as that given for $\Tgrad$.  Extracting the $\tau$ dependence indicates that, at a fixed $\Tgrad$, $\kcut \rho_i$ should scale as $(1+\tau)^{-1}$ for small $\tau$ and as $(1+\tau)^{-1/3}$ for $\tau\sim 1$.  This is confirmed in Fig.~\ref{fig:eta1_kcut_scaling}.  A plot of the variation of the growth rate spectra with $\tau$ can be found in Appendix~\ref{app:B}.

\begin{figure}
  \centering
  \subcaptionbox{}[0.51\linewidth]{%
    \includegraphics[width=\linewidth]{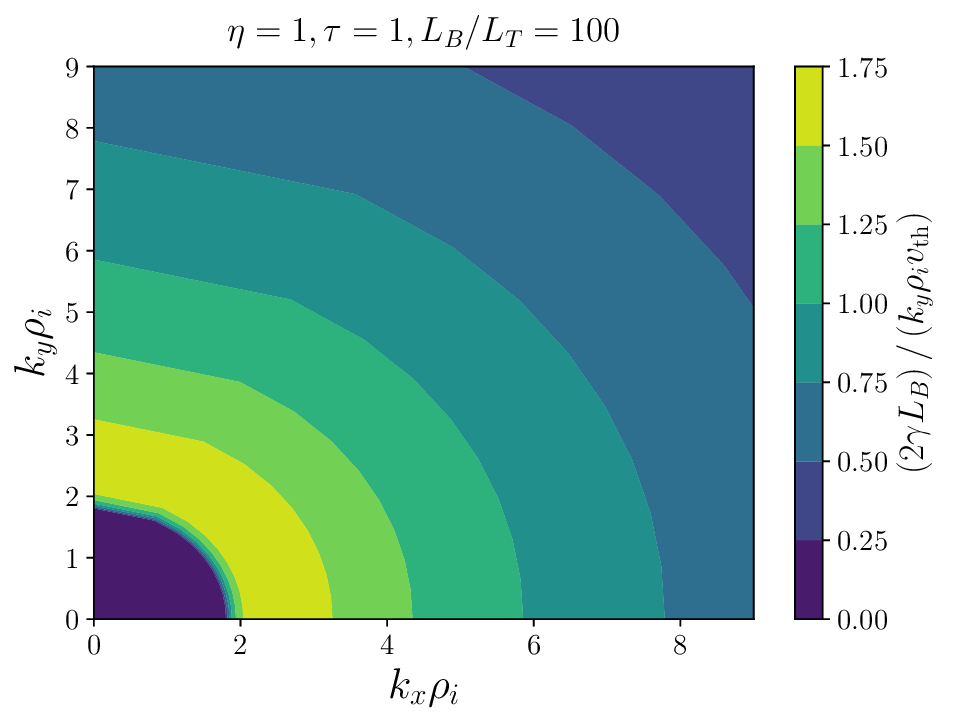}%
  }%
  \subcaptionbox{}[0.51\linewidth]{%
    \includegraphics[width=\linewidth]{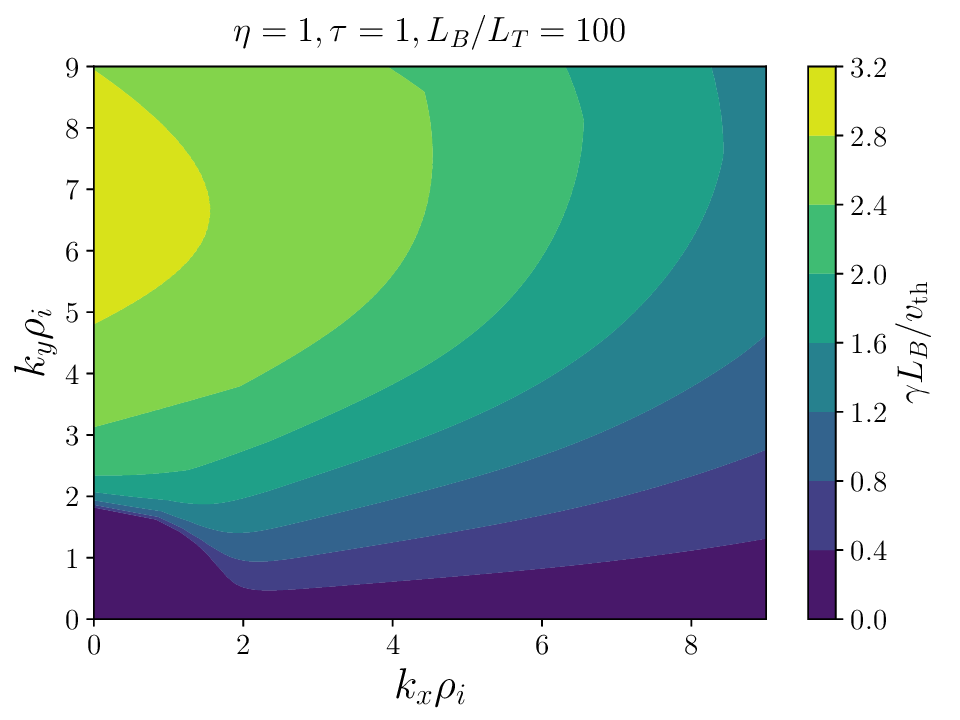}%
  }

  \caption{2-D spectral plots of the growth rate produced using the solver for $\Tgrad = 100$ and $\eta =\tau = 1$. Plot (a) is isotropic due to the presence of a normalisation factor proportional to $k_y$, while plot (b) omits the $k_y$ in its normalisation.}
  \label{fig:two2dinter}
\end{figure}

Before exploring the implications of the $\kcut$ scalings for SWITG-driven turbulence, we assess the role played by the radial wave number $k_x$. Examining the gyrokinetic dispersion relation Eq.~\eqref{eq:gk_dispersion}, one can see that the solution for $\omega/k_y$ must be isotropic in the $(k_x,k_y)$ plane.  This is because the only other appearance of $k_x$ and $k_y$ in the dispersion relation is via the $k_{\perp}$ that appears in the argument of the Bessel function $J_0$.  We demonstrate this isotropy in Figure~\ref{fig:two2dinter}, where we plot $(\gamma L_B/\vthi)(2/k_y\rhoi)$ in the $(k_x,k_y)$ plane for $\eta=\tau=1$.  Note that the instability peaks at $k_x = 0$, but it is possible for the system to be unstable even for $k_y < k_y^o$ when we allow $k_x$ to be non-zero. However, as expected from the isotropy of $\gamma/k_y$, there appears to be a circular region of $k_{\perp} \rho_i \lesssim 1$ where the system is always stable.

\section{Implications for turbulence}
\label{sec:nonlin}

We have seen in Sec.~\ref{sec:linGK} that the SWITG instability has a wavenumber cutoff that increases as the ion temperature gradient increases.  If $\eta$ is large enough for the long-wavelength ITG to be destabilised, then the turbulence driven by the SWITG is unlikely to contribute significantly to transport: Large eddies are more effective at mixing than small eddies.  However, if $\eta$ is small enough for the long-wavelength ITG to be stabilised, then SWITG-driven transport could become important.  A possible implication for the nonlinear dynamics associated with SWITG modes is that the outer scale of the turbulence will shift to shorter wavelengths as the driving gradients increase.  To see the ramifications for transport, we adopt a simple diffusive estimate for the heat flux, with the diffusion coefficient $\chi$ estimated as
\begin{equation}
    \chi \sim \frac{1}{\left(k_{\perp}^o\right)^2 \tau_{\textnormal{nl}}^o}.
\end{equation}
Here, $\tau_{\textnormal{nl}}^o$ is a measure of the characteristic nonlinear decorrelation time at the outer scale, and we have taken the characteristic radial extent of an eddy at the outer scale to be of size $1/k_{\perp}^o$, with $k_{\perp}^o$ the cutoff wavenumber for the linear SWITG mode.  Note in this estimate that we have assumed $k_x^o \sim k_{\perp}^o$.  For the remainder of this section the superscript `o' indicates a quantity evaluated at the outer scale.

We estimate the nonlinear decorrelation time at the outer scale by balancing the linear injection with the nonlinear transfer to smaller scales. To see how this works, first note from quasineutrality~\eqref{eq:qnk} that $\int d^3 v \ J_0(\bessarg) \hk \sim (e\phik/T)n$, where we assume for simplicity that $T\defeq T_i=T_e$.  Taking the velocity moment of the gyrokinetic equation~\eqref{eq:gkk} with the nonlinear term included then gives
\begin{equation}
\frac{\partial}{\partial t}\int d^3 v \ J_0(\bessarg) \hk^o \sim  \left(\int d^3v \gyroR{\boldsymbol{v}_E}\cdot\nabla h\right)_{\boldsymbol{k}^o} \defeq \frac{e\phik^o}{T}\frac{n_i}{\tau^o_{\textnormal{nl}}},
\end{equation}
where the middle term is the Fourier component of the $E\times B$ nonlinearity corresponding to the outer scale wave vector $\boldsymbol{k}^o$.  Balancing this nonlinear transfer term with the linear drive finally yields
\begin{equation}
\frac{e\phik^o}{T}\frac{n_i}{\tau^o_{\textnormal{nl}}}\sim  \Gamma_0(b^o_k) \omega^o_* \eta \frac{e\phik^o}{T}n_i,
\end{equation} 
from which we obtain $\tau_{\textnormal{nl}}^o \sim 1/(\Gamma_0(b_k^o)\omega_*^o \eta)$.  The diffusivity then scales as
\begin{equation}
\chi \sim \frac{\vthi}{L_T}\frac{\Gamma_0(b_k^o)}{k_{\perp}^o}.
\end{equation}
When $L_B/L_T$ is very large, $k_{\perp}^o\rhoi \gg 1$ so that $\Gamma_0(b_k^o) \sim 1/(k_{\perp}^o\rhoi)$ and the asymptotic scaling of Eq.~\eqref{eq:outerscale} should be valid.  In this limit, the diffusivity should \textit{decrease} with increasing temperature gradient: $\chi \propto L_T$.  This corresponds to an ion heat flux that is independent of ion temperature gradient.  For the intermediate range of $L_B/L_T$ values shown in Fig.~\ref{fig:eta1_kcut_scaling}, $k_{\perp}^o \propto L_T^{-1/3}$, but $\Gamma_0$ does not have a clear asymptotic scaling.  Consequently, no power-law scaling for $\chi$ can be inferred in this non-asymptotic regime, though the empirical scaling for $k_{\perp}^o$ suggests a weak (positive) scaling of $\chi$ with $L_T^{-1}$.

The SWITG instability we have considered is two-dimensional, but the turbulence it drives is unlikely to be.  For two different planes perpendicular to the magnetic field to remain correlated, information must propagate between them.  This provides a lower bound on the allowed $k_{\parallel}$ for turbulent structures. The critical balance conjecture~\citep{goldre,barnes2011critically,Adkins_Schekochihin_Ivanov_Roach_2022,niesPRL2025,adkins2026asymptoticscalingtheoryelectrostatic} provides a means of determining $k_{\parallel}$: The parallel correlation length of the turbulence is the distance that information can propagate before it is decorrelated by the turbulent dynamics in the plane perpendicular to the magnetic field.  In our case, the waves of interest are sound waves, and critical balance implies $k_{\parallel}\vthi\sim \taunl^{-1}$.  Using the above expression for $\taunl^o$ gives a scaling for the parallel outer scale with ion temperature gradient:
\begin{equation}
    k_{\parallel}^o \sim \frac{1}{L_T}.
\end{equation}
Thus, we anticipate that turbulent eddies driven by the SWITG mode have characteristic wavelengths both along and across the magnetic field that decrease with increasing temperature gradient.  The aspect ratio of the eddies at the outer scale, $\epsilon^o\defeq k_{\perp}^o/k_{\parallel}^o$, is predicted to decrease with increasing temperature gradient as $\epsilon^o\sim L_T^{2/3}$ until very large $\Tgrad$, when $\epsilon^o$ should become independent of $L_T$.

\section{Discussion}
\label{sec:discussion}

In this paper we have considered the linear properties of the short-wavelength ITG (SWITG) instability for plasma consisting of a single ion species and Boltzmann electrons immersed in a constant-curvature magnetic field.
When the ratio of the ion density gradient and temperature gradient scale lengths, $\eta$, is sufficiently small (cf.~\eqref{eq:etacrit} and Fig.~\ref{fig:criteta}), the long-wavelength branch of the curvature ITG mode is stabilised.  However, the SWITG mode remains unstable.  We have shown that, unlike its long-wavelength counterpart, the SWITG has a cutoff wavenumber that \textit{increases} with increasing temperature gradient: It follows the power-law scalings observed in Fig.~\ref{fig:eta1_kcut_scaling} and predicted (at large temperature gradient) by Eq.~\eqref{eq:outerscale}.  An immediate, counter-intuitive implication is that the linear growth rate of the SWITG mode at any fixed wavenumber will eventually decrease to zero as the temperature gradient increases, as illustrated in Fig.~\ref{fig:eta1_2D_growthrate}.

We showed in Sec.~\ref{sec:nonlin} that a simple diffusive estimate for the fluxes leads to scaling-law predictions for the ion heat flux and turbulence spatial and temporal scales associated with SWITG turbulence.  Adoption of the critical balance conjecture further provides scalings for the parallel extent of turbulent eddies. For moderate values of the ion temperature gradient ($\Tgrad \sim 10$), the parallel turbulence scale is predicted to decrease with increasing $\Tgrad$ faster than the perpendicular scale, and the ion heat flux $Q$ is predicted to scale weakly with ion temperature gradient as $Q \propto (\Tgrad)^{2/3}$.  For large values of the ion temperature gradient ($\Tgrad\gtrsim 100$), both the parallel and perpendicular turbulence outer scales are predicted to be inversely proportional to $\Tgrad$, and the ion heat flux is predicted to be independent of the ion temperature gradient.

It should be emphasised that the analysis we have performed is limited in a variety of ways.  Perhaps the most severe assumption we have made is that of a Boltzmann response for the electrons, which removes kinetic electron effects and electron-driven instabilities that can coexist with -- and thus modify -- the SWITG mode (cf.~\cite{chowdhuryPoP09}).  The use of a constant-curvature field removes particle trapping, magnetic shear and other physics associated with the variation of the magnetic field strength along a field line -- each of which has been found to quantitatively modify SWITG mode growth rates~\citep{gao2005, chowdhuryPoP09}.  We have ignored electromagnetic effects~\citep{gao2005, gaj2020}, radial profile shearing~\citep{chowdhuryPoP09} and numerous other factors that would be present in a realistic tokamak or stellarator plasma.  The aim of our heuristic analysis is to give an indication of how the SWITG instability and associated turbulence properties should behave as the plasma density and temperature gradients are varied.  A natural next step is to check the scaling-law predictions provided here using nonlinear gyrokinetic simulations in both constant-curvature and more realistic tokamak or stellarator geometries, which we leave to a future publication.

\section*{Acknowledgements}
This work was supported by the Engineering and Physical
Sciences Research Council (EPSRC)  [EP/R034737/1]. The authors would like to thank P. Luhadiya for helpful discussions regarding the semi-analytical solver described in this work. 
\appendix
\section{Construction of the FLR solver} \label{app:flr_solver}
We adopt the notation used in \citet{Ivanov_Adkins_2023} and follow the prescription described in their §7 to solve the resonant integral in equation \eqref{eq:gk_dispersion} numerically while incorporating FLR effects.  

Equation \eqref{eq:gk_dispersion} can be  expressed in the form
\begin{equation}
   1 + \tau + \left[\omega - \omega_* + \eta \omega_* \left(\partial_s + \partial_t + \frac{3}{2}\right) \mathcal{I}_{s,t}|_{s=t=1} \right]  = 0,
\end{equation}
where\footnote{Note that $\mathcal{I}_{s,t}$ as defined here is the 2-D simplification ($k_{\parallel} \vthi \to 0$) of $\mathcal{I}_{s,t}$ as defined in \citet{Ivanov_Adkins_2023}.} 
\begin{equation}
    \mathcal{I}_{s,t} = \frac{1}{\sqrt{\pi}} \int_{-\infty}^{\infty} \mathrm{d}u \int_0^{\infty} \mathrm{d} \mu \frac{e^{-su^2 - t\mu}}{-\omega + \omega_d (2u^2 + \mu)} J_0^2(\bessarg).
\label{eq:Idef}
\end{equation}
The factor of $J_0^2(x)$ appearing in \eqref{eq:Idef} can be represented as a convergent power series \citep{watson1922treatise} such that
\begin{equation}
    J_0^2(\bessarg) = \sum_{m = 0}^{\infty} \frac{(-1)^m (2m)!}{(m!)^4} \left(\frac{\bessarg}{2}\right)^{2m},
\end{equation}
where $\bessarg^2 = \mu (k_{\perp} \rho_i)^2$.  We thus have 
\begin{equation}
    \mathcal{I}_{s,t} = \sum_{m = 0}^{\infty} \frac{(2m)!}{(m!)^4} b_k^m \partial^m_t I_{s,t}, \label{eq:3.4}
\end{equation}
where $b_k \equiv (k_{\perp}\rho_i)^2/2$, and $I_{s,t}$ is the resonant integral in the drift-kinetic limit, i.e.,
\begin{equation}
    I_{s,t} = \frac{1}{\sqrt{\pi}} \int_{-\infty}^{\infty} \mathrm{d}u \int_0^{\infty} \mathrm{d} \mu \frac{e^{-su^2 - t\mu}}{-\omega + \omega_d (2u^2 + \mu)}.
\label{eq:Iab}
\end{equation}

From \eqref{eq:Iab}, we see that $I_{1,1}$ is the same as equation $\eqref{eq:gk_dispersion}$ in the drift-kinetic limit and can be expressed in terms of the plasma dispersion function~\citep{biglari1989toroidal,Zocco_Xanthopoulos_Doerk_Connor_Helander_2018} as 
\begin{equation}
    I_{1,1} = -\frac{1}{2\omega_d} Z^2(\sqrt{\hat{\omega}}),  \quad \quad \hat{\omega} = \frac{\omega}{2\omega_d}.
\end{equation}
The plasma dispersion function \citep{1961pdf..book.....F} is defined to be 
\begin{equation}
    Z(x) = \frac{1}{\sqrt{\pi}} \int_{-\infty}^{\infty} \mathrm{d}y \frac{e^{-y^2}}{y - x}.
\end{equation}

\citet{Ivanov_Adkins_2023} also provide analytical expressions for the derivatives of $I_{s,t}$ in terms of $I_{s,t}$. Hence, if the $m^{\mathrm{th}}$ derivative of $I_{s,t}$ can be determined recursively in terms of $I_{s,t}$ and then evaluated at $s = t = 1$, we can subsequently solve equation \eqref{eq:gk_dispersion} numerically. This cannot be done in general but is possible for the 2-D case that we are concerned with. This is what we will now describe.

The derivative of $I_{s,t}$, derived in Appendix B of \citet{Ivanov_Adkins_2023}, is
\begin{equation}
    \partial_t I_{s,t} = \frac{1}{s-2t} \left[I_{s,t} (-2\hat{\omega} (s-2t) + 1) -\frac{2}{\omega_d} \sqrt{\hat{\omega}} Z_s(\sqrt{\hat{\omega}}) -\frac{\sqrt{s}}{t \omega_d}\right],
\end{equation}
where $Z_s(x) = Z(\sqrt{s}x)$. Generalisation to the $m$th derivative gives
\begin{align}
    \partial^m_t I_{s,t}  = &\frac{1}{s-2t} \left[(2m-1 -2\hat{\omega} (s-2t)) \partial^{m-1}_t I_{s,t} \right. \nonumber \\  + & \left. 4\hat{\omega}(m-1) \partial^{m-2}_t I_{s,t} +  
    \left(\frac{-1}{t}\right)^m \frac{\sqrt{s}}{\omega_d} (m-1)! \right]. \label{eq:3.9}
\end{align}
Setting $s = t = 1$ allows us to compute the power series in equation \eqref{eq:3.4}. To compute ($\partial_s + \partial_t) \partial^m_t I_{s,t}$, we note that $(\partial_s + \partial_t)f (s,t)|_{t=s} = \partial_s f(s, t = s)$.

Hence, we set $t = s$ in equation \eqref{eq:3.9} and differentiate with respect to $s$:
\begin{align}
    \partial_s (\partial_t^m I_{s,t}|_{t=s}) & = \frac{-1}{s} \bigg[\partial^m_t I_{s,t}|_{t=s} + 2\hat{\omega} \partial^{m-1}_t I_{s,t}|_{t=s} + \left(2m-1 + 2\hat{\omega} s \right)\partial_s(\partial_t^{m-1}I_{s,t}|_{t=s}) \nonumber \\
    & + 4\hat{\omega}(m-1) \partial_s(\partial_t^{m-2}I_{s,t}|_{t=s}) + \frac{(-1)^{m+1}}{s^{m+1/2}\omega_d}(m-1)!\left(m-\frac{1}{2}\right)\bigg]. \label{eq:3.10}
\end{align}

When evaluated at $s = t = 1$, equations \eqref{eq:3.9} and \eqref{eq:3.10} represent the recursion relations that allow us to solve \eqref{eq:gk_dispersion} numerically. Closure of the recursive process requires explicit specification of the left-hand sides of equations~\eqref{eq:3.9} and~\eqref{eq:3.10} for $m = 0$ and $m= 1$. For \eqref{eq:3.9}, these quantities are
\begin{equation}
    I_{1,1} = -\frac{1}{2\omega_d} Z^2(\sqrt{\hat{\omega}}) \quad \quad \partial_t I_{s,t}|_{t = s =1} = \frac{1}{2\omega_d} \left[ Z^2(\sqrt{\hat{\omega}})(1 + 2\hat{\omega}) + 4\sqrt{\hat{\omega}}Z(\hat{\omega}) + 2\right],
\end{equation}
and for equation \eqref{eq:3.10},
\begin{align}
    \partial_s I_{s,s} |_{s=1} & = \frac{1}{4\omega_d} Z^2(\sqrt{\hat{\omega}}) + \frac{1}{\omega_d} \sqrt{\hat{\omega}}Z(\sqrt{\hat{\omega}}) \left[1 + \sqrt{\hat{\omega}}Z(\sqrt{\hat{\omega}})\right], \nonumber \\
    \partial_s (\partial_t I_{s,t}|_{t=s})|_{s=1} & = -\partial_s I_{s,s}|_{s=1} (2\hat{\omega} + 1) + \frac{1}{2\omega_d} Z^2(\sqrt{\hat{\omega}}) - \frac{2}{\omega_d} \sqrt{\hat{\omega}} Z(\sqrt{\hat{\omega}}) \nonumber \\
    & - \frac{2}{\omega_d} \hat{\omega} \left[ 1 + \sqrt{\hat{\omega}} Z(\sqrt{\hat{\omega}})\right] -\frac{3}{2 \omega_d} .
\end{align}

Using these recursion relations and a quick numerical implementation of the plasma dispersion function such as from \citet{fad}, one can solve the linearised gyrokinetic equation \eqref{eq:gk_dispersion} to arbitrarily high $k_\perp \rho_i$ using a root finding algorithm. This is provided that one uses enough terms in the expansion for the Bessel function such that the expansion is a sufficiently good approximation to the actual value at a given $k_\perp \rho_i$. The number of terms required increases rapidly with the value of $k_\perp \rho_i$.  We used $m=60$ for $k_\perp \rho_i = 6$, $m = 150$ for $k_{\perp}\rhoi =8$ (cf. Fig.~\ref{fig:eta_infty_1Dcut_growthrate}), $m = 250$ for $k_\perp \rho_i = 10$ (cf. Fig.~\ref{fig:eta1_1Dcut_growthrates}), $m = 450$ for $k_\perp \rho_i = 12$ (cf. Fig.~\ref{fig:eta_infty_2D_growthrate}), and $m = 650$ for $k_\perp \rho_i = 13$ (cf. Fig.~\ref{fig:two2dinter}).

\section{Additional growth rate spectra} \label{app:B}

Here, we present additional 1D cuts of the growth rate spectra discussed in Section~\ref{sec:linGK}, highlighting key features of the instability.  

\begin{figure}
    \centering
    \includegraphics[width = 0.8\linewidth ]{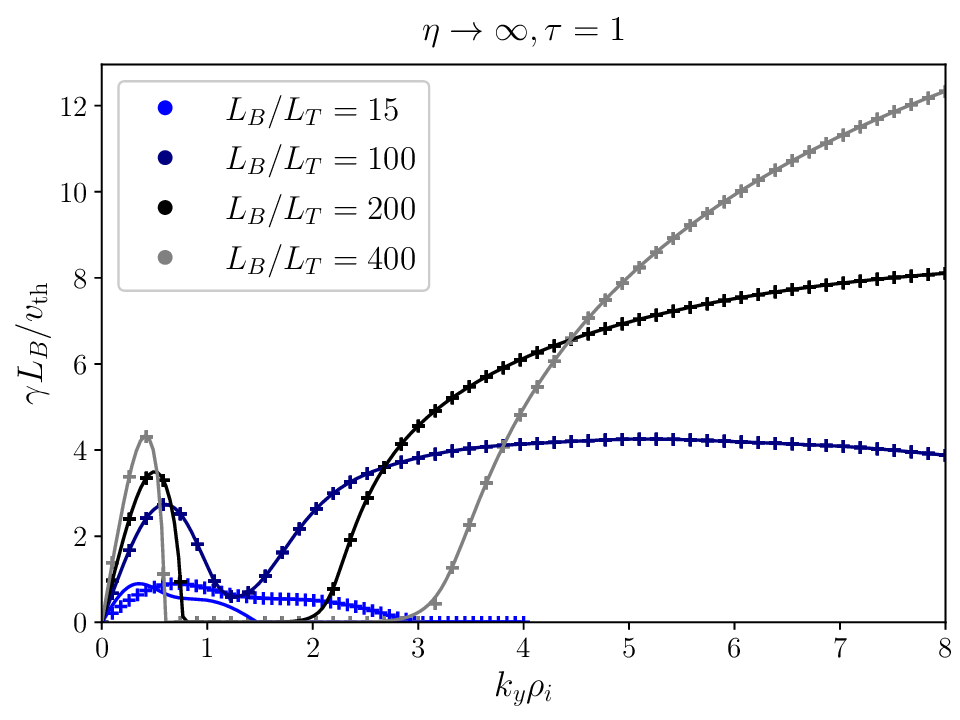}
    \caption{Plots of the growth rate spectra made using \texttt{stella} (crosses) and the semi-analytical solver (lines) with $\eta \to \infty$ and $\Tgrad = 15$, 100, 200, and 400.}
    \label{fig:3}
\end{figure}

We first show in Figure \ref{fig:3} the growth rate spectra in the $\eta\rightarrow\infty$ limit for several different values of $\Tgrad$.  Note that the SWITG cutoff wavenumber increases with $\Tgrad$: at large $\Tgrad$ values, the long-wavelength and short-wavelength ITG spectra are distinct, and they merge at more modest values of $\Tgrad$, giving the characteristic double-bump structure seen in previous studies.

\begin{figure}
    \centering
    \includegraphics[width = 0.8\linewidth]{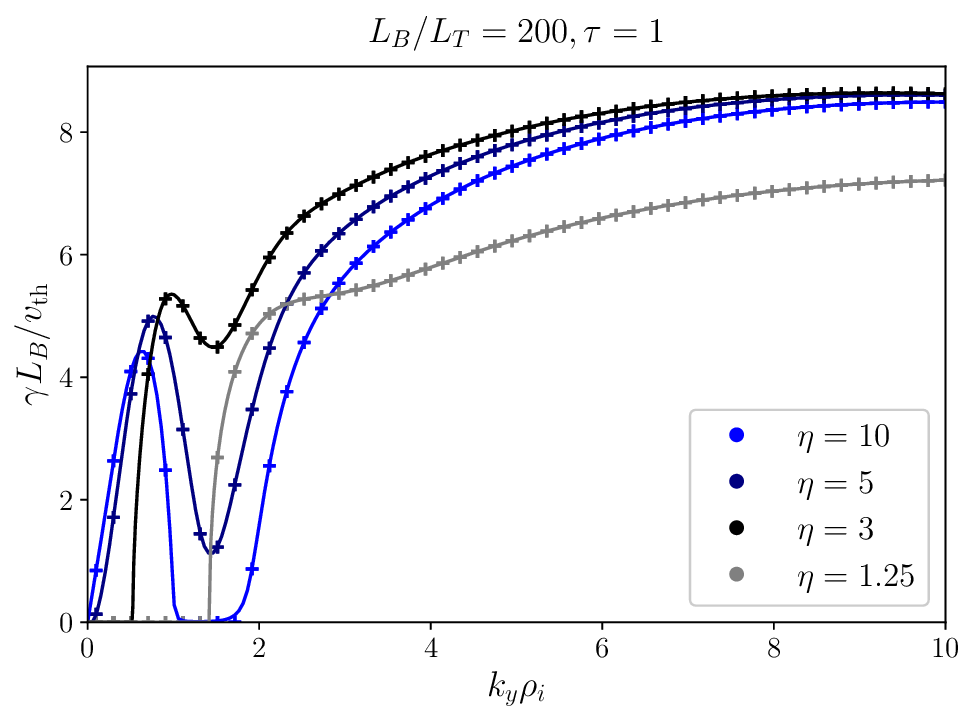}
    \caption{Plots of the growth rate spectra made using \texttt{stella} (crosses) and the semi-analytical solver (lines) for different values of the density gradient at fixed $L_B /L_T = 200$, $\tau = 1$.}
    \label{fig:8}
\end{figure}

\begin{figure}
    \centering
    \includegraphics[width = 0.8\linewidth]{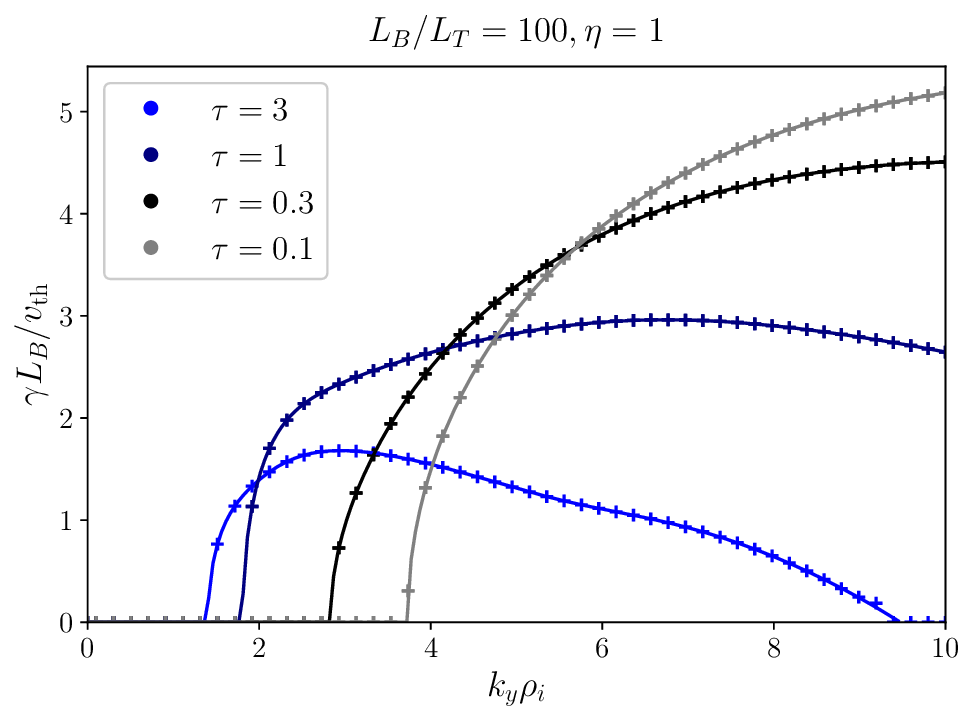}
    \caption{Plots of the growth rate spectra made using \texttt{stella} (crosses) and the semi-analytical solver (lines) for $\Tgrad = 100, \eta = 1$, and $\tau = 3$, 1, 0.3, and 0.1.  Decreasing $\tau$ has a similar effect to increasing $\Tgrad$; i.e., the cutoff wavenumber of the instability is pushed to larger values.}
    \label{fig:7}
\end{figure}

The variation of the growth rate spectrum with density gradient is illustrated in Figure~\ref{fig:8}.  The data is taken from simulations with fixed $\Tgrad=200$, $\tau=1$ and with $\eta$ varying from 1.25 to 10.  The main effect of an increasing density gradient is to stabilise the long-wavelength ITG, leading to a clear cutoff at long-wavelengths for the SWITG.  There is also a modest, non-monotonic modification to the SWITG growth rates.

Finally, the effect of varying the ion-electron temperature ratio $\tau$ is captured in Figure \ref{fig:7}, in which the growth rate spectra are plotted for fixed $\eta=1$, $\Tgrad=100$ and $\tau$ varying from 0.1 to 3.  The cutoff wavenumber for the SWITG increases as $\tau$ decreases, in line with the predictions of the heuristic fluid model presented in Section~\ref{sec:fluid_model}.

\bibliographystyle{jpp}

\bibliography{bibliography}

\end{document}